\documentclass[eqsecnum,floats,aps,prd,floatfix,titlepage,tightenlines]{revtex4} 

\usepackage{graphicx}
\usepackage{graphics}
\usepackage{bm}
\usepackage{amssymb}
\usepackage{amsmath}
\usepackage{mathrsfs}

\begin{document}
\title{ Vacuum Radiation Pressure Fluctuations on Electrons}
     \author{L. H. Ford}
     \affiliation{Institute of Cosmology, Department of Physics and Astronomy, Tufts University, Medford, Massachusetts 02155, USA}
     
     \begin{abstract}
     This paper is a continuation of a  study of the properties and applications of quantum stress tensor fluctuations.  Here we treat the vacuum fluctuations
     of the electromagnetic energy-momentum flux operator which as been averaged in space and time. The probability distribution of these fluctuations
     depends upon the details of this averaging and may allow fluctuations very large compared to the variance. The possibility of detecting their effects on
     electrons will be considered. The averaging of the flux operator will arise from the interaction of an electron with a wave packet containing real photons,  
     The vacuum radiation pressure fluctuations can exert a force on the electron in any direction, in contrast to the effect of scattering by real photons.
     Some numerical estimates of the effect will be given.
     
     \end{abstract}

\maketitle

\baselineskip=14pt

\section{Introduction}
\label{sec:intro}

This paper will deal with quantum fluctuations of radiation pressure and their possible observable effects. The vacuum fluctuations of stress tensor operators
has been a topic of several investigations in recent years~\cite{FewsterFordRoman:2010}-\cite{WSF23}.  
 A key result which has emerged is the the probability distribution for these fluctuations is very sensitive
to the details of how they are measured. On a formal level, a quadratic operator, such as a stress tensor component, must be averaged in time in order to have
a well defined probability distribution. Physically, this averaging is linked to the details of the measurement process. The rate of decay of the Fourier transform
of the averaging functions with increasing frequency determines the asymptotic form of the probability distribution. Typically, this form is an exponential of a small
fractional power. As a result, the probability of large fluctuations can be orders of magnitude larger than would have been predicted by a Gaussian distribution.
Some possible observable effects of these large fluctuations might include enhancement of the quantum tunneling rates~\cite{HF17,HF22}, 
or light scattering by zero point density fluctuations in a liquid~\cite{FS09,WF20} . 
 
In the present paper, a different process will be addressed: the effects of quantum radiation pressure fluctuations on the motion of electrons. A discussion of radiation
pressure fluctuations on atoms was given in Ref.~\cite{F21}.
A key feature of the
probability distribution for vacuum radiation pressure fluctuations is being symmetric; the pressure is equally to occur in any direction.  In contrast, the radiation pressure
due to real photons exerts a force in the direction in which the photons are traveling. Here a model will be presented in which space and time averaging of the 
electromagnetic momentum flux operator is produced by a localized wave packet containing real photons. The scattering of real photons by an electron can give the electron
linear momentum in the direction of motion of the wave packet. However, the vacuum radiation pressure fluctuations can potentially contribute momentum in the opposite
direction. If this contribution can be observed, this could constitute observation of vacuum radiation pressure fluctuations.

The outline of this paper is as follows: Section~\ref{sec:rad-press-op} will review selected aspects of the quantum radiation pressure operator and its 
fluctuations. In particular,  Sect.~\ref{sec:large-flucts} will deal with large fluctuations.  Section~\ref{sec:elecron} will present a model for the interaction of
both real photons and vacuum fluctuations with an electron. Some numerical estimated of the magnitude of the vacuum effects will be given. The results
of the paper will be summarized in Sect.~IV.

Units in which $c = \hbar =1$ will be used unless otherwise noted.

\section{The Radiation Pressure Operator}
\label{sec:rad-press-op}

The electromagnetic momentum flux in the $z$-direction is
\begin{equation}
T^{tz}(t, \mathbf{x})    = (\mathbf{E} \times \mathbf{B})^z = E^x\, B^y -E^y\, B^x  \,,
\end{equation}
where $\mathbf{E}$ and $\mathbf{B}$ are the quantized electric and magnetic field operators, respectively. Recall that in units where
$\hbar = c = 1$, the electromagnetic momentum flux operator is also the energy flux operator, the Poynting vector.

\subsection{Coherent States and the Classical Limit}
\label{sec:coherent state}

A classical electromagnetic wave may be considered to be a highly excited coherent state. For the case of a single excited mode,
such a state may be defined as an eigenstate of the photon annihilation operator for this mode:
\begin{equation}
 a\,|z \rangle = z \, |z \rangle
 \end{equation}
 where $z$ is an arbitrary complex number,  The mean number of photons in this state is
 \begin{equation}
 \langle a^\dagger\, a \rangle = |z|^2 \, ,
 \end{equation}
  and the variance in this number is
 \begin{equation}
  \langle (a^\dagger\, a)^2 \rangle -  \langle a^\dagger\, a \rangle^2 = |z|^2 \,.
 \end{equation}

If $|z| \gg 1$, then the fractional fluctuations are small, and the expectation values of the  electric and magnetic field operators approximate
classical solutions of Maxwell's equations. However, fluctuations around these mean values are always present on some level, and can produce physically observable effects including radiation pressure fluctuations on a mirror. 

The mean momentum flux in a coherent state, $\langle z|  \, T^{tz} \, |z \rangle$, is the classical radiation pressure. In quantum theory, this 
pressure may be viewed as originating from the momentum carried by the individual photons in the coherent state. Similarly, quantum fluctuations
around the mean value may be viewed as arising from fluctuations in photon number~\cite{Caves80,Caves81,WF01}. 
Consider the case of a coherent state for a traveling wave mode
moving in the $+z$ direction., in which case $\langle z|  \, T^{tz} \, |z \rangle > 0$.   Fluctuations in the number of photons passing by per unit time lead
to fluctuations around this mean value, but will never change its sign. In this picture, the minimum value of the radiation pressure is zero, and this value
is only reached when no photons arrive.

The situation is quite different in quantum field theory. Here the vacuum state can be a source of rich phenomenology, including the Casimir effect 
and the Lamb shift.

\subsection{Vacuum Fluctuations and Spacetime Averages}
\label{sec:t-ave}

 The vacuum fluctuations of the radiation pressure operator are well-defined only if it has been averaged in time. 
Let $S^z$ be the momentum flux sampled in both time and space with averaging functions $f(t)$ and $g(\mathbf{x})$
 \begin{equation}
 S^z = \int_{-\infty}^\infty T^{t z}(t,{\bf x})\, f(t)\, g(\mathbf{x}) dt\, d^3 {x} \,.
 \label{eq:Sz-def}
 \end{equation}
Note that the operators $T^{t z}$, and hence $S^z$, are automatically
 normal ordered, as   $\langle0|T^{t z}|0\rangle =  \langle0|S^z|0\rangle  =0$.  We assume that $f(t)$ and  $g(\mathbf{x})$
 are infinitely differentiable, non-negative functions which are normalized so that
\begin{equation}
\int_{-\infty}^\infty  f(t)\, dt =   \int  g(\mathbf{x}) \,  d^3 {x} \,  =  1 \,.
\end{equation}

Here $f(t)$ and  $g(\mathbf{x})$ are interpreted as describing the effect of a physical measurement of the averaged momentum flux. 
Although spatial averaging is not essential for $S^z$ to be well defined, here we assume it is present as a part of the measurement.
 Furthermore, we require that $f = g=0$ outside of a finite spacetime region, as a physical measurement should occur in such a region.
  As a consequence, $f(t)$ and $g(\mathbf{x})$ cannot be  analytic functions. Such functions which are nonzero in a finite interval are 
 referred to as having compact support. Define their Fourier transforms by
\begin{equation}
\hat{f}(\omega) =\int_{-\infty}^\infty dt \,{\rm e}^{-i\omega t} f(t)\,,
\label{eq:Fourier}
\end{equation}
and
\begin{equation}
\hat{g}(\mathbf{k}) = \int d^3 x \,  {\rm e}^{i \mathbf{k}\cdot \mathbf{x}}\, g(\mathbf{x}) \,.
\end{equation} 
 Infinitely differentiable 
but compactly supported functions must have a Fourier transform which decays faster than any power of $\omega$ as $\omega \rightarrow \infty$, but more slowly than an exponential.

A class of such compactly supported functions is described in Sect.~II of Ref. ~\cite{FF15}. For these functions, $\hat{f}(\omega)$  decays
as an exponential of a fractional power:
\begin{equation}
\hat{f}(\omega) \sim \gamma \, {\rm e}^{- \beta \, (\tau \omega)^\alpha } \,.
\label{eq:fhat-asy}
\end{equation}
Here $\tau$ is the characteristic temporal duration of $f(t)$, and the above form holds when $ \tau \omega \gg 1$. This asymptotic form 
depends upon the constants $0 < \alpha < 1$, $\beta >0$, and $\gamma$. The value of $\alpha$ is especially crucial, as it determines the 
magnitude of the vacuum fluctuations of $S^z$. Smaller values of $\alpha$ cause $\hat{f}(\omega)$ to fall more slowly as $\omega$ increases,
leading to larger fluctuations. We expect $\hat{g}(\mathbf{k})$ to have a similar asymptotic form for large $k$, except with $\tau$ replaced
by the characteristic spatial sampling scale, $\ell$.

Suppose that the electric and magnetic field operators are expanded in terms of plane wave modes, with photon creation and  annihilation
operator coefficients, $ a^\dagger_j$ and $ a_j $, where $j =(\mathbf{k}, \lambda)$ is a mode label describing a photon's wavevector
$\mathbf{k}$ and polarization $\lambda$. In this basis, the averaged flux operator  $S^z$  is represented as
\begin{equation}
S^z = \sum_{i\, j} (A_{i j}\, a^\dagger_i \,a_j + B_{i j}\, a_i \,a_j + 
B^*_{i j} \, a^\dagger_i \,a^\dagger_j ) \,.
\label{eq:Sz2}
\end{equation}
Here
\begin{equation}
A_{i j} = \frac{ \sqrt{\omega_i \omega_j} }{3 V}\, \hat{f}(\omega_i -\omega_j ) \, \hat{g}(\mathbf{k}_i-\mathbf{k}_j)\,,
\label{eq:A}
\end{equation}
and
\begin{equation}
B_{i j} =    \frac{ \sqrt{\omega_i \omega_j} }{3 V}\ \, \hat{f}(\omega_i +\omega_j )\,   \hat{g}(\mathbf{k}_i+ \mathbf{k}_j)\,,
\label{eq:B}
\end{equation}
where $V$ is a quantization volume and $\omega_j  = |\mathbf{k_j}|$ is the mode angular frequency. Note that Eqs,~\eqref{eq:A} and
~\eqref{eq:B} have the same form as the corresponding relations for the operator $:\dot{\varphi}^2:$, given in Ref.~\cite{FF15} , where $\dot{\varphi}$
is the time derivative of a massless scalar field.  The origin of the numerical factors in Eqs.~\eqref{eq:A} and \eqref{eq:B} is explained in 
Sect IIB of Ref.~\cite{HF17}.

Recall that time averaging with $\tau > 0 $ is essential for a quadratic operator such as $S^z$, but space averaging is not. The averaged operator
is still well defined in the limit where we average in time at a single spatial point, in which case  $g(\mathbf{x}) \rightarrow \delta(\mathbf{x})$ and
$\hat{g}(\mathbf{k})  \rightarrow 1$.

\subsection{The Moments and Eigenstates of $S^z$}
\label{sec:eigenstates}

For the case of a massless scalar field in two spacetime dimensions, it is possible to give explicit exact expressions for the probability distribution
of vacuum flux fluctuations for selected choices of the temporal sampling function~\cite{FF24}. No such exact results are known in four dimensions,
but there are at least two approaches for approximately determining the probability distribution $P(x)$ for stress tensor fluctuations. Here 
\begin{equation}
 x = \tau^4 \, S^z
 \label{eq:x-def}
 \end{equation}
is a dimensionless measure of the flux. The first approach involves calculation of the moments: 
\begin{equation}
 \mu_n =  \langle0|(S^z)^n|0\rangle \, ,
 \label{eq:mu-n}
 \end{equation}
and was used in Refs.~\cite{FF15,FF2020}. The rate of growth of $\mu_n$ as $n$ increases may be used to infer asymptotic form, 
Eq.~\eqref{eq:tail}. More specifically, the form of  $P(x)$ near a given value of $x$ is determined by the $\mu_n$ where \cite{FF2020}.
\begin{equation}
 n \approx x^{\alpha/3}
 \label{eq:n-worldline}
 \end{equation}
 in the world line limit, and
 \begin{equation}
  n \approx x^{\alpha}
  \label{eq:n-STA}
 \end{equation}
 in the spacetime averaged limit.
 
Unfortunately, the moments approach suffers from the 
ambiguity that the moments grow too rapidly to satisfy the Hamburger moment criterion, which is the condition under which the set
of moments $\{ \mu_n \}$ uniquely determine $P(x)$. However, the moments do determine the averaged features of the distribution.
When two distinct distributions possess the same moments, they typically differ by an oscillatory function which averages to zero.
Fortunately there is an alternative method which is free of this ambiguity. This is diagonalization of an operator of the form of 
Eq.~\eqref{eq:Sz2} by a Bogolubov transformation. This involves a linear transformation of the photon creation and annihilation
operators $a_j^\dagger$ and $a_j$ to a new basis with operators $b_k^\dagger$ and $b_k$ in which $S^z$ takes the diagonal form 
\begin{equation}
S^z = \sum_{k} \lambda_{k}\, b^\dagger_k \,b_k \;+ \, C  \,,
\label{eq:Sz}
\end{equation}
where $C$ and the $\lambda_{k}$ are constants. The eigenstates of $S^z$  are number eigenstates  in the new basis, $| n_k \rangle_b$, where
$b^\dagger_k \,b_k | n_k \rangle_b =   n_k\, | n_k \rangle_b$. The corresponding eigenvalues are $n_k \,  \lambda_{k} +C$. The probability
distribution in the physical vacuum, $|0\rangle$, is found from the probability amplitude, $\langle 0 |  n_k \rangle_b$, to find the $k$ th eigenvalue in a
measurement on $|0\rangle$. The $b$-mode eigenstates are multi-mode squeezed vacuum states in the $a$-mode basis, which is the
basis of physical photon states.
In practice, the diagonalization of  $S^z$ needs to be performed numerically. This was done for the operator $:\dot{\varphi}^2:$ in 
Refs.~\cite{SFF18,WSF21}, with results which agree well with the moments approach.

For our present purposes, the eigenstates of $S^z$ are the outcomes of a physical measurement of the spacetime averaged momentum flux,
and hence of the radiation pressure on an electron, which will be the topic of the Sect.~\ref {sec:elecron}.

\subsection{Typical Fluctuations}
\label{sec:typical-flucts}

The typical vacuum flux fluctuations have a variance given by the second moment
\begin{equation}
 \mu_2 =  \langle0|(S^z)^2|0\rangle = 2 \sum_{ij} B_{ij}\, B^*_{ij} \,,
 \label{eq:mu2}
 \end{equation}
and a root-mean-square value given by $S^z_{\rm rms} = \sqrt{\mu_2} = x_{\rm rms}/\tau^4$. Here $x_{\rm rms}$ is a dimensionless
constant whose magnitude depends upon the specific choice of the sampling function. The explicit form of $\mu_2$ in the 
$V \rightarrow \infty$ may be found from Eq.~\eqref{eq:B} to be
\begin{equation}
  \mu_2 =  \frac{1}{288\, \pi^6} \, \int d^3 k_1 \, d^3 k_2 \, \omega_1 \, \omega_2 \; |\hat{f}(\omega_i +\omega_j )|^2 \,   |\hat{g}(\mathbf{k}_i+ \mathbf{k}_j)|^2\,.
   \label{eq:mu2-STA}
 \end{equation}

Now consider the worldline limit, where $\tau \gg \ell$. In this case we have
\begin{equation}
 \mu_2 =\frac{1}{18\, \pi^4 }\, \int_0^\infty d\omega_1\, d\omega_2 \, (\omega_1 \, \omega_2 )^3 \, |\hat{f}(\omega_1+\omega_2)|^2 \,.
  \label{eq:mu2-worldline}
 \end{equation}
As $\alpha$ decreases,   $\mu_2$ and hence  $x_{\rm rms}$  increase, due to an increasing contribution from high frequency modes,
 as may be seen from Eq.~\eqref{eq:fhat-asy}. 
An explicit example of an $\hat{f}(\omega)$ with $ \alpha = 1/2$ is the function $\hat{L}(\omega)$ in Sect. IIB of Ref.~\cite{FF15}, which 
is approximated by the function $\hat{h}_{\rm fit}(\omega)$ in Appendix A of Ref.~\cite{FF2020}.  In this case, $x_{\rm rms} \approx 2.5$. 

Quantum fluctuations of both linear fields and stress tensors exhibit subtle correlations  and anti-correlations. These may be described by a 
correlation function, which for the flux operator can take the form
\begin{equation}
 C(t,t') =  \langle0|S^z(t) \,  S^z(t')  |0\rangle\,.
 \end{equation}
Here $S^z(t)$ and $S^z(t')$ denote flux operators averaged over different spacetime regions localized near times $t$ and $t'$, If $C(t,t') > 0$, then
a flux measurement near $t$ is positively correlated with one near $t'$. Similarly, $C(t,t') < 0$ implies anti-correlation. The correlations of
vacuum energy density fluctuations is discussed in Refs.~\cite{FR05,FR07}, for example, 
where it was argued that these fluctuations tend to be anti-correlated.
Thus negative energy density tends to be either proceeded of followed by positive energy density. We can expect a similar anti-correlation in
vacuum energy flux fluctuations. Note that $C(t,t')$ describes the correlations of typical fluctuations, and that  $C(t,t)  = \mu_2(t)$, the variance of 
the variance of the flux at $t$.

\subsection{Large Fluctuations}
\label{sec:large-flucts}

Here we summarize previous results on the asymptotic probability distribution for stress tensor fluctuations.  Recall that $x$, defined in
Eq.~\eqref{eq:x-def},
is a dimensionless measure of a momentum flux fluctuation. Then $ x \approx x_{\rm rms} = O(1)$ represents a typical fluctuation, and
$x \gg 1$ is a large fluctuation. For large $x$, the probability distribution $P(x)$ has the form
\begin{equation}
P(x)  \sim c_0 \,x^b \, {\rm e}^{-a x^c} \,,
\label{eq:tail}
\end{equation}
where the constants $a$, $b$, $c$, and $c_0$ depend upon the choice of sampling function. 

Often we are more interested in the probability of fluctuations larger than a given magnitude. This is described by the complementary 
cumulative distribution
\begin{equation}
  P_>(x) = \int_x^\infty P(y)\, dy \,.
  \end{equation} 
For $x \gg 1$, the asymptotic form is
\begin{equation}
  P_>(x) \approx    \frac{c_0}{a\, c}\; x^{1+b-c}\, {\rm e}^{-a x^c} \,.
  \label{eq:cumP}
 \end{equation} 
Hence, both $P(x)$ and $ P_>(x)$ decrease with the same exponential of a fractional power.

 A natural question which arises is, how to describe the correlations of large fluctuations? This is still an unanswered question. The usual
 correlation function approach using  functions as $C(t,t')$ describes the correlations of typical, not large, fluctuations. Here an alternative 
 approach will be suggested. Recall that the large fluctuations are determined by the large moments, as described by Eqs.~$\eqref{eq:n-worldline}$
 and $\eqref{eq:n-STA}$. This motivates the definition of {\it generalized correlation functions} of the form
 \begin{equation}
 C_{m,n}(t,t') =  \langle 0|(S^z)^m(t) \,  (S^z)^n(t')  |0\rangle\,.
 \end{equation}
 Thus, $C_{11}$ is the usual correlation function, and $C_{nn}$ is the correlation function for the operator $(S^z)^n$. It will be a topic of future 
 research  to study applications to the correlations of large fluctuations.

\subsubsection{Worldline Averaging}
\label{sec:WLA}

In the limit when the spatial sampling scale is sufficiently small, we are essentially averaging the stress tensor along a timelike worldline.
In this case, the exponent $c$ in the asymptotic form of the probability distribution is given by
\begin{equation}
c = \frac{\alpha}{3}  \,.
  \label{eq:c-worldline}
 \end{equation}
 Because $\alpha < 1$, this implies a slowly decreasing tail with $c < 1/3$, with an enhanced probability for large fluctuations. The criterion 
 for the validity of the worldline approximation is
 \begin{equation}
 x \alt \left(\frac{\tau}{\ell}\right)^3 \,.
 \label{eq:worldline}
 \end{equation}

\subsubsection{Spacetime Averaging}
\label{sec:SLA}

When
\begin{equation}
 x \agt \left(\frac{\tau}{\ell}\right)^3 \, ,
 \end{equation}
the worldlne approximation no longer holds, and the effects of spatial averaging become important. In this case, the exponent $c$
becomes
\begin{equation}
c \approx \alpha\,.
  \label{eq:c-STA}
 \end{equation}
This behavior has been confirmed by the results of numerical diagonalization~\cite{WSF21}. As $x$ increases for fixed $\tau/\ell \gg 1$,
the probability distribution $P(x)$ makes a smooth transition from the form described by Eq.~\eqref{eq:c-worldline} to that of  
 Eq.~\eqref{eq:c-STA}. The enhanced rate of decrease of $P(x)$ reflects the role of spatial averaging in suppressing large vacuum
 fluctuations.
 
 Because $\alpha < 1$, the probability distribution is still decreasing more slowly than an exponential function, and much more slowly
 than the Gaussian distribution which describes random processes. This is a reflection of the fact that quantum fluctuations are highly
 correlated.

\section{Effects on an Electron}
\label{sec:elecron}

\subsection{Electron Momentum Fluctuations}
\label{sec:p-flucts}

Now we take up the possibility of observing radiation pressure fluctuations by their effects on the motion of an electron. Assume that the 
electron is initially at rest in the laboratory frame, and is then subjected to radiation pressure  with a duration of  $\tau$ and a
magnitude of $|S^z| = x/\tau^4$. Here we assume that the photons, whether real or virtual, producing the radiation pressure have energies
of the order of the electron rest mass $m$ or smaller. In this case the photon-electron interaction may be treated as Thompson scattering,
for which the cross section is
\begin{equation}
 \sigma_T = \frac{8 \pi \, \alpha_{fs}^2}{3\, m^2} \,,
 \label{eq:Thompson}
 \end{equation}
where $\alpha_{fs} \approx 1/137$ is the fine structure constant. The characteristic change in the electron's linear momentum is
\begin{equation}
 \Delta p \approx S_z
 \, \tau \, \sigma_T \,.
 \end{equation}
If the electron's motion in the laboratory frame remains non-relativistic, this corresponds to a change in speed of order
\begin{equation}
  \Delta v \approx \frac{\Delta p}{m} = \frac{8 \pi \, \alpha_{fs}^2}{3\, (m\, \tau)^3} \, x \,, 
 \end{equation}
or
\begin{equation}
   \Delta v \approx \frac{1.5 \times 10^{-4}}{ (m\, \tau)^3} \, x \,.
   \label{eq:delta-v}
 \end{equation}

The probability distribution for the resulting modification of the electron's momentum is given by Eq.~\eqref{eq:tail} for $x \gg 1$, where
$x$ becomes a function of $\Delta p$:
\begin{equation}
 x(\Delta p) =     \frac{3\, m^2\, \tau^3 }{ 8 \pi \, \alpha_{fs}^2} \,.
\label{eq:delta-p}
\end{equation}
The resulting function $P(\Delta p)$ is an exponential function which depends upon a negative power of the coupling constant
$ \alpha_{fs}$, and is hence a non-analytic function. Such a function cannot arise in any finite order of perturbation theory,

\subsection{Need for Spatial Averaging}
\label{sec:Delta ell}

If we assume that the dimensions of the electron wave packet dictate the spatial averaging scale, then we need to enquire about
possible transverse spreading of this wave packet. If the initial transverse dimension is of order $\ell$, this implies a transverse
momentum of order
\begin{equation}
 p_T \approx \frac{1}{\ell} \,.
 \end{equation}
In a time $\tau$, this will lead to  transverse spreading of the order of
\begin{equation}
 \delta \ell \approx \frac{p_T}{m} \, \tau \,.
 \end{equation}
If we require that
\begin{equation}
  \delta \ell  \alt \ell \,,
 \end{equation}
then we have
\begin{equation}
 \frac{\tau}{\ell} \alt m \, \ell\,.
 \end{equation}
We may rewrite Eq.~\eqref{eq:delta-v} as
\begin{equation}
   \Delta v \approx \frac{1.5 \times 10^{-4}}{ (m\, \ell)^3} \, \left(\frac{\ell}{\tau}\right)^3\; x \,.
   \label{eq:delta-v1}
 \end{equation}
In the worldline approximation, where Eq.~\eqref{eq:worldline} holds, we have
\begin{equation}
   \Delta v \alt \frac{1.5 \times 10^{-4}}{ (m\, \ell)^3} \,.
   \label{eq:delta-v2}
 \end{equation}
 If we expect that $\ell$ is large compared to the electron Compton wavelength, $m\, \ell \agt 1$, this places a very strong constraint
 on the magnitude of $ \Delta v$ which can be achieved in the worldline approximation. For this reason, we turn to the case where
 spatial averaging is important.
 
\subsection{Probability Estimates}
\label{sec:estimates}

Here we wish to make some rough estimates of the probability for a stress tensor fluctuation of a given magnitude using results 
obtained in Refs.~\cite{FF2020} and \cite{WSF21}. These references consider the case of the operator $:\dot{\varphi}^2:$, and we 
assume that the probability of a large fluctuation of this operator is of the same order as those of $S^z$. We further assume that
the spatial sampling function, $g(\mathbf{x})$, is spherically symmetric, and that its Fourier transform,
$\hat{g}(\mathbf{k}) = \hat{g}(k)$, decays  for large $k$ as does $\hat{f}(\omega)$ in Eq.~\eqref{eq:fhat-asy}, but with $\tau$
replaced by $\ell$. As described in Sect IIB of Ref.~\cite{FF2020}, $\hat{g}(k)$ is determined by a one-dimensional function 
$\hat{h}(\omega)$. The asymptotic spacetime averaged distribution in the spacetime averaged limit is found to have the form
\begin{equation}
 P(x) \propto  {\rm e}^{-(1+\sqrt{2 s})\, \sqrt{x/B}} \,,
 \label{eq:STA-P}
 \end{equation}
for $s = \ell/\tau \agt 1$, where $\alpha = 1/2$.
The constant $B$ in this case is
\begin{equation}
 B = \frac{f(0)}{ 8 \pi  \, |h''(0)| \, s^2 }\,.
 \end{equation}
 Note that as $s$ decreases, $B$ increases, and $P(x)$ falls more slowly with increasing $x$. This reflects the fact that for fixed $x$ and decreasing $s$,
$P(x)$ will eventually transition to the worldline form given by Eq.~\eqref{eq:c-worldline}..
 The proportionality constant in Eq.~\eqref{eq:STA-P} is not uniquely determined by the moments approach used in Ref.~\cite{FF2020}, but could be found 
 by the numerical diagonalization approach in Refs.~\cite{SFF18,WSF21}. For the purposes of an order of magnitude estimate,  we will assume that this 
 proportionality constant is of order one.
 
 The probability of a fluctuation of $x$ or greater is now found from Eq.~\eqref{eq:cumP} to be of order
 \begin{equation}
 P_> \approx     \frac{2 \sqrt{B}}{1+\sqrt{2 s}}\, x^{1/2} \, P(x)\,.
 \label{eq:P-approx}
 \end{equation}

If we use the specific sampling functions described in  Ref.~\cite{FF2020}, where
\begin{equation}
 f(0) \approx 1.5 \qquad {\rm and} \qquad  |h''(0)| \approx 0.076 \,,
 \end{equation}
 we find $B\approx 1.0/s^2$ and
\begin{equation}
 P(x) \approx  {\rm e}^{-, (1 +\sqrt{ 2 s})\, s \, \sqrt{x}}
  \label{eq:Pg-approx}
 \end{equation}   
for the case $\alpha = 1/2$ and $b \approx 0$.
This result appears to be supported by the numerical calculations in Ref.~\cite{WSF21} in the upper panel of Fig.~2, where $s \approx 0.14$.
 
 Now we consider a few numerical examples:
 
 \subsubsection{$m \tau =100$ , $\ell =0.1 \, \tau$ , and $x = 10^4$}
 
 Here we find
 \begin{equation}
 \Delta v \approx 1.5  \times 10^{-6}
 \end{equation}
 and
 \begin{equation}
 P_>  \approx 2.3 \times 10^{-4} \,.
 \end{equation}
 
 \subsubsection{$m \tau =1$ , $\ell = \tau$ , and $x = 10$}

In this case, where we are at the limit of the non-relativistic approximation, the results are
 \begin{equation}
 \Delta v \approx 1.5 \times 10^{-3}
 \end{equation}
 and
 \begin{equation}
 P_>  \approx 1.3 \times 10^{-3} \,.
 \end{equation}

Here we are considering a pulse of photons in the gamma ray energy range. Such pulses might be generated by the back scattering 
of optical frequency photons from high energy electrons~\cite{Milburn63,Milburn65}.

\subsection{What constitutes a measurement of  $S^z$ ?}
\label{sec:measurement}

The mathematical definition of the averaged momentum flux operator is given in  Eq.~\eqref{eq:Sz-def}. However, we would like to 
link the sampling functions $f(t)$ and $g(\mathbf{x})$  to a physical measurement. One option seems to let them be determined
by the envelope function of a wave packet mode function. Let $g(\mathbf{x})$ be proportional to the envelope function at a fixed 
time $t$, and let $f(t)$ be proportional to the envelope function at a fixed point in space as a function of time. Suppose that the 
electromagnetic field is in a coherent state of this mode.  The scattering of the real photon in this state by an electron not only
produce radiation pressure, but can also serve to measure  changes in motion of the electron.   Potentially, this can include changes 
due to vacuum radiation pressure fluctuations.  If the real photons in the wave packet are moving in the $+z$ direction, a single
photon-electron scattering event produces a recoil in the same direction, as discussed in the Appendix. In the presence of several electrons,
an electron can recoil in the opposite direction due to back scattered photons, but is effect will depend upon the electron density.
The vacuum fluctuations are equally likely to
produce a force in the opposite direction, and hence electron recoil in the $-z$ direction.

A related question is how many photons need to be in the coherent state for the scattering of the photons to constitute a measurement
of  $S^z$? If we need a large mean number of photons, $|z| \gg 1$, then the scattering by real photons may mask the effects of the
vacuum fluctuations. An alternative may be that repeated measurements by wave packets with a relative small number of photons
may be effective in measuring $S^z$, and hence the vacuum radiation pressure fluctuations. If this is the case, then a mean number
of photons which is small compared to one may suffice if the number of repeated measurements is large. In this case, vacuum fluctuations 
could dominate over the effects of photon number fluctuations.

\section{Summary}
\label{sec:final}

A model for the detection of quantum radiation pressure fluctuations has been presented. The quantum radiation pressure operator is averaged
in space and time by the interaction of a localized photon wave packet with an electron. This averaged operator will have vacuum fluctuations
which are equally likely be in any direction and to have an asymptotic probability distribution which decays as an exponential of a fractional power, 
leading to the possibility of large fluctuations. If the quantum  state of the electromagnetic field is a coherent of real photons traveling in the $+z$-direction, 
the expectation value of the radiation pressure on an electron will also be in this direction, and is due to the effect of photon-electron scattering.
However, there will be two sources of fluctuations in this pressure: (1) Fluctuations in the number of photons in the coherent state~\cite{WF01}. These cause 
variations in the magnitude of the pressure, but cannot change its sign. This is due to the fact discussed in the Appendix that an electron at rest
scattering with a photon moving in the $+z$-direction will recoil in this direction. (2) Vacuum fluctuations of the radiation pressure operator, which
are equally likely to impart momentum in either direction.  If the operator averaging function may be taken to be the wave packet envelope function,
then the probability of large fluctuations depends upon the asymptotic Fourier transform of the envelope function. and hence the asymptotic power
spectrum of the photon pulse.

Some numerical estimates for the probability of the vacuum fluctuations required for various electron speeds are given Sect.~\ref{sec:estimates}.
At the limit of the non-relativistic approximation we see that $\Delta v \approx 10^{-3}$ could occur with a probability of the order of a few times $10^{-3}$.
This might be observable if the vacuum fluctuation effect can be dis-entangled from the effects of scattering by real photons. This might be 
possible if the mean number of photons can be made small enough. This is topic for further work.  As noted in Sec.~\ref{sec:p-flucts}, the effects of
large vacuum radiation pressure fluctuations on the state of the electron is a non-perturbative effect.

Another future topic will be extension of the present model beyond the non-relativistic approximation. This approximation arises in part from the 
use of the Thomson cross section, Eq.~\eqref{eq:Thompson}, to describe the scattering  of virtual photons by an electron. However, the Compton
cross section decreases slowly (as a logarithm) as the photon energy increases above the electron rest mass energy. This suggest that the
non-relativistic approximation may be a good order of magnitude estimate at these higher energies. 

Other topics for further work include a better understanding of the correlations between large fluctuations discussed in Sect.~\ref{sec:large-flucts},
and of the non-perturbative correction to the electron's quantum state discussed in Sect.~\ref{sec:p-flucts}.

\begin{acknowledgments} 
I would like to thank Chris Fewster and Ken Olum for useful conversations.
This work was supported in part  by the National Science Foundation under Grant PHY-2207903.
\end{acknowledgments}

 \appendix
 \section{Electron-Photon Scattering Kinematics}
 
\begin{figure}[htbp]
\includegraphics[scale=0.7]{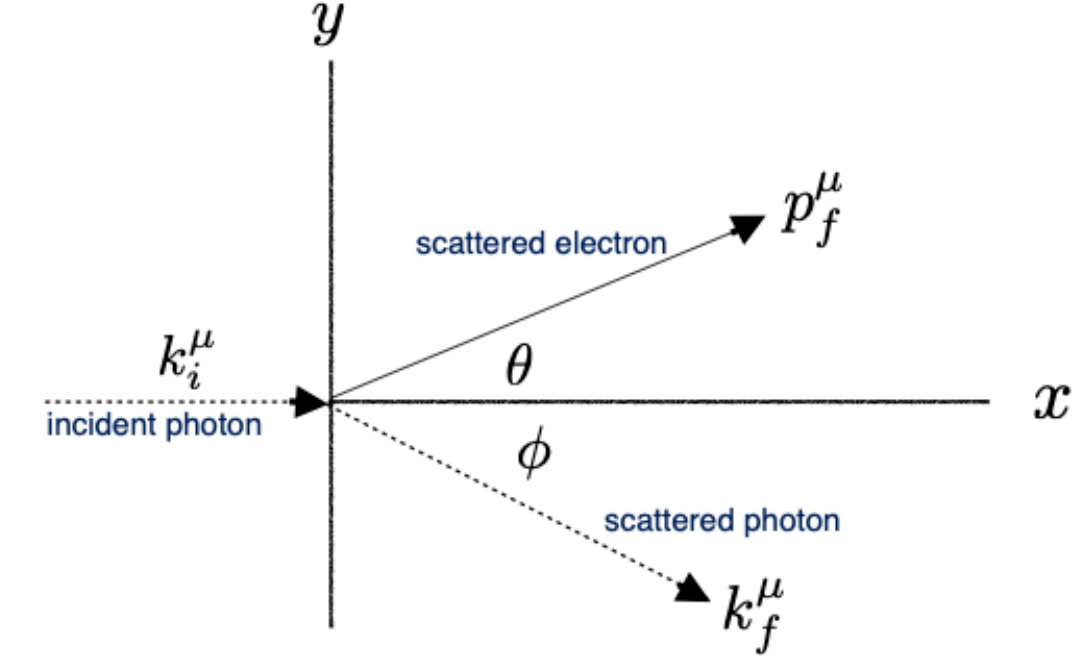}
\caption{A photon with initial four-momentum $k^\mu_i$, moving in the $+x$-direction, collides with an electron at rest and scatters at angle $\phi$ 
and four-momentum $k^\mu_f$.  The electron recoils at angle $\theta$ and four-momentum $p^\mu_f$.} 
\label{fig:scattering}
\end{figure}

 Here we review the relativistic kinematics of an electron-photon collision. 
 We also derive the key result that an electron which is initially at rest in the laboratory  frame cannot recoil into the direction from which the photon came. 
 In this frame, an initial photon  moving in the $+x$ direction with energy $\omega_0$ collides 
 with an electron at rest, as  illustrated in Fig.~\ref{fig:scattering}. 
 
 The electron recoils with a speed $v$ at an angle $\theta$, and the photon with energy $\omega$ at an angle
 $\phi$, as shown. The initial four-momenta of the electron and photon are
 \begin{equation}
 p^\mu_i = (m,0,0,0)
 \end{equation}
 and
 \begin{equation}
 k^\mu_i= \omega_0 (1,1,0,0)\,,
 \end{equation}
respectively. The corresponding final momenta are
 \begin{equation}
  p^\mu_f= m\, \gamma (1,v\, \cos \theta,  v\, \sin \theta,0)
 \end{equation}
 and
 \begin{equation}
  k^\mu_f = \omega (1, \cos \phi, \sin \phi,0)\,,
 \end{equation}
 where $\gamma = (1-v^2)^{-\frac{1}{2}}$, or $v =   (1-\gamma^2)^{\frac{1}{2}}/\gamma$.

Conservation of energy and momentum, $p^\mu_i +  k^\mu_i= p^\mu_f+  k^\mu_f$ lead to
\begin{equation}
 \omega = 1 - \omega_0 -\gamma \,,
 \label{eq:omega}
 \end{equation}
 \begin{equation}
  \omega _0 =  (1- \gamma^2)^{\frac{1}{2}}\, \cos \theta + \omega\, \cos \phi\,,
 \end{equation}
 and
 \begin{equation}
  (1- \gamma^2)^{\frac{1}{2}}\,   \sin \theta =  =  \omega \, \sin \phi \,.
 \end{equation}
 Here units in which $m = 1$ have been adopted. Use of \eqref{eq:omega}  to eliminate $\omega$ leads to
  \begin{equation}
  \omega _0 =  (1- \gamma^2)^{\frac{1}{2}}\, \cos \theta +   (1 - \omega_0 -\gamma)\, \cos \phi\,,
 \end{equation}
 and
 \begin{equation}
  (1- \gamma^2)^{\frac{1}{2}}\,   \sin \theta = (1 - \omega_0 -\gamma) \, \sin \phi \,.
 \end{equation}
 Finally, use of $\cos^2 \phi + \sin^2 \phi =1$ leads
 to the result
\begin{equation}
 \cos \theta = \sqrt{\frac{\gamma - 1}{\gamma + 1}} \; \frac{\omega_0 + 1}{\omega_0} \,.
 \end{equation}
 
 Because $\gamma \geq 1$, this implies $\cos \theta \geq 0$, and hence $\theta \leq \pi/2$. This is our key result, that the electron cannot back scatter
 into the direction from which the photon came. There is a simple intuitive explanation: Energy conservation requires $\omega < \omega_0$. The final photon energy must be less that the initial photon energy to provide kinetic energy to the electron. As a result, the final $+x$ component of the photon's momentum
 will be less than its initial value. This requires that the electron recoil in the  $+x$ direction to conserve momentum.

\end{document}